\documentstyle[12pt,aps,graphicx]{revtex}

\begin{document}

\title{
Constraints on Non-Singular Cosmological Models with Quadratic Lagrangians
}

\author{
R. Colistete Jr.\thanks{Email address: {\tt coliste@ccr.jussieu.fr}}
}

\address{
L.P.T.L., Universit\'e Paris VI\\Tour 22-12, 4\`eme \'etage,
Bo\^{\i}te 142, 4 place Jussieu, 75005 Paris - France
}

\maketitle

\begin{abstract}

We consider the generalized set of theories of gravitation whose Lagrangians
contain the term $R^{2}$ : $L=\sqrt{-g}(R+\beta R^{2})$. Inserting the RW
metric with an imposed non-singular and inflationary behaviour of the scale
factor $a(t)$, and using a arbitrary perfect fluid, we study the properties
of $\rho $ and $p$ in this context. By requiring the positivity of the
energy density , as well as real and finite velocity of sound, we can obtain
the range of values of $\beta $ that ensure the inflationary behaviour and
absence of singularity.

\end{abstract}

\section{Introduction}

There are many motivations to study quadratic-order curvature Lagrangians in
cosmology: quantum corrections\cite{Lanczos,Pais} and renormalization of
divergences,\cite{DeWitt,Hooft,Stelle} effective actions of superstring
theory,\cite{Shapiro} inflation without scalar fields,\cite{Starobinsky} etc.

The aim of the present work is to explore the consequences of adding
quadratic-order curvature terms to the Einstein action. From the beginning,
a specified non-singular and inflationary scale factor and the
Robertson-Walker metric are imposed, and the unknown energy-momentum tensor
is then analysed.

The only energy-momentum tensor $T_{\mu \nu }$ choice in this work will be
an arbitrary perfect fluid described by the energy density $\rho $ and the
isotropic pressure $p$. Just a couple of simple energy conditions will be
used here, 
\begin{equation}
\rho >0\ , \quad \quad (\rho +p)>0\ ,
\label{EnergyCond}
\end{equation}
and additionally the adiabatic speed of sound, $v_{s}^{2}=(\partial
p/\partial \rho )$, must obey 
\begin{equation}
0\leq v_{s}^{2}\leq 1\ .
\label{VsCond}
\end{equation}

For each value of the spatial curvature $k$, the form of $\rho $ and $p$ are
obtained as functions of the scale factor $a(t)$ and other constants. The
energy (\ref{EnergyCond}) and velocity of sound (\ref{VsCond}) conditions
can give constraints on the constant $\beta $ associated with the
quadratic-order curvature terms. It is also possible to compare the
violations of the conditions (\ref{EnergyCond},\ref{VsCond}) with and
without the quadratic-order curvature terms, which gives some insight on the
consequences of using quadratic Lagrangians.

\section{The modified Friedmann equation}

Considering the fact\cite{Lanczos} that the quantity $\sqrt{-g}(R_{\mu \nu
\rho \sigma }R^{\mu \nu \rho \sigma }-4R_{\mu \nu }R^{\mu \nu }+R^{2})$,
known as Gauss-Bonnet term, is a total divergence in four-dimensional
spacetime, the $R_{\mu \nu \rho \sigma }R^{\mu \nu \rho \sigma }$ term can
be expressed in terms of $R^{2}$ and $R_{\mu \nu }R^{\mu \nu }$, so the
Riemann tensor is not necessary to account for the curvature square
corrections.

Representing a general quadratic order curvature Lagrangian (but without the
cosmological constant), let us take the Lagrangian 
\begin{equation}
{\it L}=\sqrt{-g}(R+\beta _{1}R^{2}+\beta _{2}R_{\mu \nu }R^{\mu \nu })\ .
\label{LagMod}
\end{equation}
Calculating the variation of this action with respect to the metric $g^{\mu
\nu }$, the modified Einstein equations are 
\begin{equation}
G_{\mu \nu }+\beta _{1}W_{\mu \nu }^{(1)}+\beta _{2}W_{\mu \nu}^{(2)}=%
\kappa\,T_{\mu \nu }\ .
\label{EinsteinMod}
\end{equation}

We will use the Robertson-Walker (RW) metric 
\begin{equation}
ds^{2}=-{\rm d}t^{2}+a(t)^{2}\biggr[\biggr(\frac{1}{1-k\ r^{2}}\biggl){\rm d}%
r^{2}+r^{2}\biggr({\rm d}\theta ^{2}+\sin ^{2}(\theta ){\rm d}\varphi ^{2}%
\biggl)\biggl]\ ,
\label{RWmetric}
\end{equation}
where the spatial curvature $k$ takes the values $0$, $+1$,$-1$
(corresponding to flat, closed and open spatial sections). In the case of
this spherically symmetric homogeneous metric, the Weyl tensor has null
components. So the identity $C_{\mu \nu \rho \sigma }C^{\mu \nu \rho \sigma
}=R_{\mu \nu \rho \sigma }R^{\mu \nu \rho \sigma }-2R_{\mu \nu }R^{\mu \nu }+%
\frac{1}{3}R^{2}$ is equal to zero, yielding another equation that now
allows us to write $R_{\mu \nu }R^{\mu \nu }$ in terms of $R^{2}$ :
\begin{equation}
{\it L}=\sqrt{-g}\left[ R+\left( \beta _{1}+\frac{\beta _{2}}{3}\right) R^{2}%
\right] =\sqrt{-g}(R+\beta R^{2})\ .
\label{LagMod2}
\end{equation}

With the use of a perfect fluid, the modified Friedmann third-order
differential equation is obtained : 
\begin{eqnarray}
\frac{{a^{\prime }(t)}^{2}+k}{{a(t)}^{2}}-\frac{\kappa \,\rho (t)}{3}-\frac{%
18\,\beta \,}{{a(t)}^{4}}\left\{ k^{2}-3\,{a^{\prime }(t)}^{4}-{a(t)}^{2}\,{%
a^{\prime \prime }(t)}^{2}-\right. &
\label{eqFrMod} \\
\left. 2\,{a^{\prime }(t)}^{2}\,\left[ k-a(t)\,a^{\prime \prime }(t)\right]
+2\,{a(t)}^{2}\,a^{\prime }(t)\,a^{\prime \prime \prime }(t)\right\} &
=0\ ,
\nonumber
\end{eqnarray}
where $\beta =(\beta _{1}+\frac{\beta _{2}}{3})$ different from zero means
that the quadratic-order curvature corrections are present.

With the RW metric (\ref{RWmetric}), the energy-momentum conservation
equations yield 
\begin{equation}
\rho {^{\prime }(a)=}\frac{{-3(}\rho +p)}{a}\ ,
\label{EnergyCons}
\end{equation}
so that $p(a)$ can be calculated using $\rho (a)$.

\section{Non-singular scale factors with inflationary \newline
behaviours}

Non-singular and inflationary scale factors for the three values of spatial
curvature $k$ are studied. They show inflationary behaviour matching the
classical radiation dominated behaviour.

\bigskip

{\bf I) The }$k=0${\bf \ case :}

\medskip

\begin{figure}[h]
\begin{center}
\includegraphics[scale=0.80]{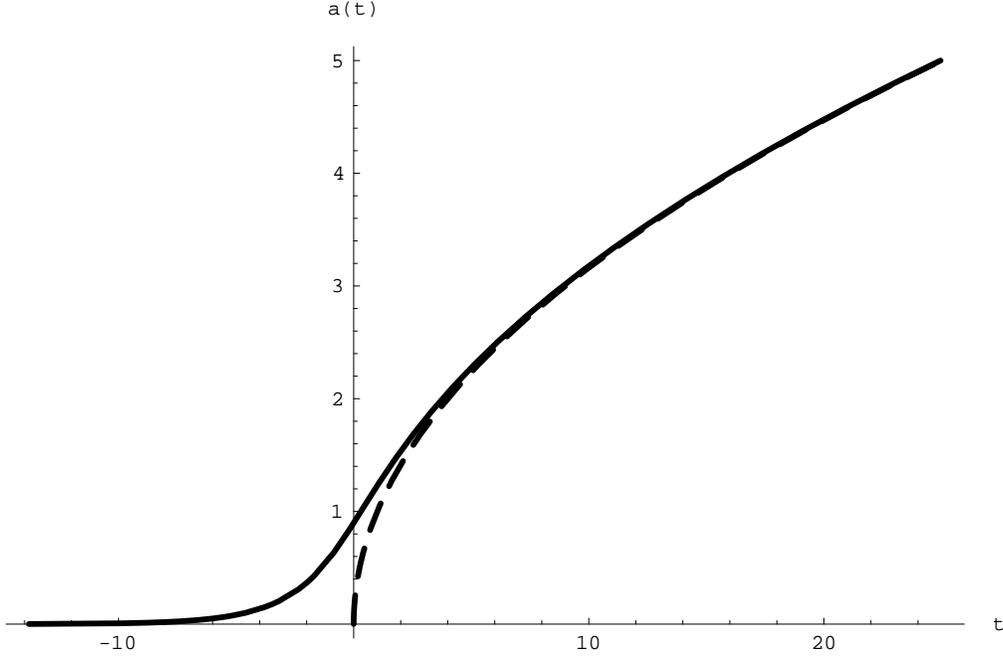}
\caption{$a(t) \times t$ for the case $k=0$, with $a_0 = \alpha = 1$. The
solid line is the non-singular scale factor with inflationary behaviour;
the dashed line is the known scale factor, $a(t)=a_{0}\sqrt{t}$, for radiation
era in FRW cosmology.}
\label{figa}
\end{center}
\end{figure}

The following equation\cite{Kerner}
\begin{equation}
t=\frac{a(t)^{2}}{{a_{0}}^{2}}+\alpha \,\log \left[ \frac{a(t)^{2}}{a(t)^{2}+%
{a_{0}}^{2}}\right]\ ,
\label{eqamodk0}
\end{equation}
allows to obtain a scale factor $a(t),$ which for $\alpha =0$ (or $a>>a_{0}$%
) yields $a(t)=a_{0}\sqrt{t}$ (the known solution for radiation dominated
era in\ FRW cosmology with $k=0$), and for $t\rightarrow -\infty $ (or $%
a<<a_{0}$) gives $a(t)=a_{0}\exp (t/2\alpha )$. Figure \ref{figa} shows this
behaviour compared to $a(t)=a_{0}\sqrt{t}$.

It is possible to derive Eq. (\ref{eqamodk0}) in respect to $t$ and write
\begin{equation}
{a^{\prime }(t)=\frac{a(t)\,{a_{0}}^{2}\,[a(t)^{2}+{a_{0}}^{2}]}{%
2\,[a(t)^{4}+a(t)^{2}\,{a_{0}}^{2}+{a_{0}}^{4}\,\alpha ]}}\ ,
\label{eqdadtk0}
\end{equation}
deriving this Eq. (\ref{eqdadtk0}), the second derivative can be obtained, 
\begin{equation}
a^{\prime \prime }(t)=\frac{a(t)\,{a_{0}}^{8}[\,a(t)^{2}+{a_{0}}%
^{2}][3\,a(t)^{2}+{a_{0}}^{2}]\,\alpha -a(t)^{3}\,{a_{0}}^{4}\,[{a(t)^{2}+{%
a_{0}}^{2}]}^{3}}{4\,[{a(t)^{4}+a(t)^{2}\,{a_{0}}^{2}+{a_{0}}^{4}\,\alpha ]}%
^{3}}\ ,
\label{eqd2adt2k0}
\end{equation}
and deriving this Eq. (\ref{eqd2adt2k0}), the third derivative is given by 
\begin{eqnarray}
a^{\prime \prime \prime }(t)& =&\frac{a(t)\,{a_{0}}^{6}\,[a(t)^{2}+{a_{0}}%
^{2}]\,}{8\,[{a(t)^{4}+a(t)^{2}\,{a_{0}}^{2}+{a_{0}}^{4}\,\alpha ]}^{5}}%
\left\{ 3\,a(t)^{4}\,[{a(t)^{2}+{a_{0}}^{2}]}^{4}- \right.\label{eqd3adt3k0} \\
& &2\,a(t)^{2}\,{a_{0}}^{4}\,[a(t)^{2}+{a_{0}}^{2}]%
[\,15\,a(t)^{4}+15\,a(t)^{2}\,{a_{0}}^{2}+4\,{a_{0}}^{4}]\,\alpha +\nonumber\\
& & \left.{a_{0}}^{8}\,[15\,a(t)^{4}+12\,a(t)^{2}\,{a_{0}}^{2}+{a_{0}}^{4}]\,%
{\alpha }^{2}\right\}\ .\nonumber
\end{eqnarray}

The scale factor $a(t)$ and its derivatives are smooth functions of $t$.

\bigskip

{\bf II) The }$k=+1${\bf \ case :}

\medskip

Generalizing the equation Eq. (\ref{eqamodk0}), 
\begin{equation}
t=\left( 1-\sqrt{1-\frac{a(t)^{2}}{{a_{0}}^{2}}}\right) t_{0}+\alpha \,%
\log \left[ \frac{a(t)^{2}}{a(t)^{2}+{a_{0}}^{2}}\right] \ ,
\label{eqamodk1}
\end{equation}
allows to obtain $a(t),$ which for $\alpha =0$ (or $a\approx a_{0}$) yields $%
a(t)=a_{0}\sqrt{1-(1-t/t_{0})^{2}}$ (the known solution for radiation
dominated era in\ FRW cosmology with $k=+1$), and for $t\rightarrow -\infty $
(or $a<<a_{0}$) gives $a(t)=a_{0}\exp (t/2\alpha )$.

As in the $k=0$, it is possible to derive Eq. (\ref{eqamodk1}) in respect to 
$t$ and write
\begin{equation}
{a^{\prime }(t)=\frac{a(t){a_{0}}\,\sqrt{a{_{0}}^{2}-a(t)^{2}}\,[a(t)^{2}+{%
a_{0}}^{2}]}{a(t)^{4}t_{0}+a(t)^{2}\,{a_{0}}^{2}t_{0}+2{a_{0}}^{3}\sqrt{{%
a_{0}}^{2}-a(t)^{2}}\,\alpha }}\ ,
\label{eqdadtk1}
\end{equation}
and the second and third derivatives can also be obtained without problems.
All these functions are smooth.

\bigskip

{\bf III) The }$k=-1${\bf \ case :}

\medskip

Also generalizing the equation Eq. (\ref{eqamodk0}), 
\begin{equation}
t=\left( -1+\sqrt{1+\frac{a(t)^{2}}{{a_{0}}^{2}}}\right) t_{0}+\alpha \,\log %
\left[ \frac{a(t)^{2}}{a(t)^{2}+{a_{0}}^{2}}\right] \ ,
\label{eqamodkm1}
\end{equation}
allows to obtain $a(t),$ which for $\alpha =0$ (or $a>>a_{0}$) yields $%
a(t)=a_{0}\sqrt{-1+(1+t/t_{0})^{2}}$ (the known solution for radiation
dominated era in\ FRW cosmology with $k=-1$), and for $t\rightarrow -\infty $
(or $a<<a_{0}$) gives $a(t)=a_{0}\exp (t/2\alpha )$.

As in the $k=0$, it is possible to derive Eq. (\ref{eqamodkm1}) in respect
to $t$ and write
\begin{equation}
{a^{\prime }(t)=\frac{a(t){a_{0}}\,\,[a(t)^{2}+{a_{0}}^{2}]^{3/2}}{%
a(t)^{4}t_{0}+a(t)^{2}\,{a_{0}}^{2}t_{0}+2{a_{0}}^{3}\sqrt{{a_{0}}%
^{2}+a(t)^{2}}\,\alpha }}\ ,  \label{eqdadtkm1}
\end{equation}
and the second and third derivatives can also be easily obtained. All these
derivatives are smooth functions.

\section{Analysis of $\protect\rho $ and $p$}

Instead of trying to obtain solutions for $a(t)$ of the modified Friedmann
equation, Eq. (\ref{eqFrMod}), which can only be solved numerically, another
way of analysis will be chosen. Replacing the derivatives of $a(t)$, then
the energy density $\rho (t)$ can be written as $\rho (a)$, also depending
on the constants $k$ and $\beta $ and the constants of the definition of $%
a\left( t\right) $.

\bigskip

{\bf I) The }$k=0${\bf \ case :}

\medskip

The energy density is given by 
\begin{eqnarray}
\rho (a)&=&\frac{3\,{a_{0}}^{4}\,{\left( a^{2}+{a_{0}}^{2}\right) }^{2}}{4\,{%
\left( a^{4}+a^{2}\,{a_{0}}^{2}+{a_{0}}^{4}\,\alpha \right) }^{2}\,\kappa }+%
\frac{27\,a^{2}\,{a_{0}}^{12}\,{\left( a^{2}+{a_{0}}^{2}\right) }%
^{2}\,\,\alpha \,\beta }{2\,{\left( a^{4}+a^{2}\,{a_{0}}^{2}+{a_{0}}%
^{4}\,\alpha \right) }^{6}\,\kappa }\times
\label{rhok0} \\
& &\left[ 14\,a^{6}+29\,a^{4}\,{a_{0}}^{2}+2\,a^{2}\,{a_{0}}^{4}\,\left(
10-3\,\alpha \right) -5\,{a_{0}}^{6}\,\left( -1+\alpha \right) \right] \ ,
\nonumber
\end{eqnarray}
and a thorough analysis shows that $\rho \geq 0$ if $\alpha $ and $%
\beta $ are inside the region 
\begin{equation}
Q_{1}=\{0\leq \alpha <1,\beta >\beta _{-}\}\cup \{\alpha >1,\beta
_{-}<\beta <\beta _{+}\}\ ,
\label{Q1k0}
\end{equation}
where $\beta _{-}$ and $\beta _{+}$ are functions of $\alpha $. The fact of
adding the non-linear terms in the Lagrangian is represented by
$\beta \neq 0$, and does not force $\rho $ to be negative.

\begin{figure}[h]
\begin{center}
\includegraphics[scale=0.80]{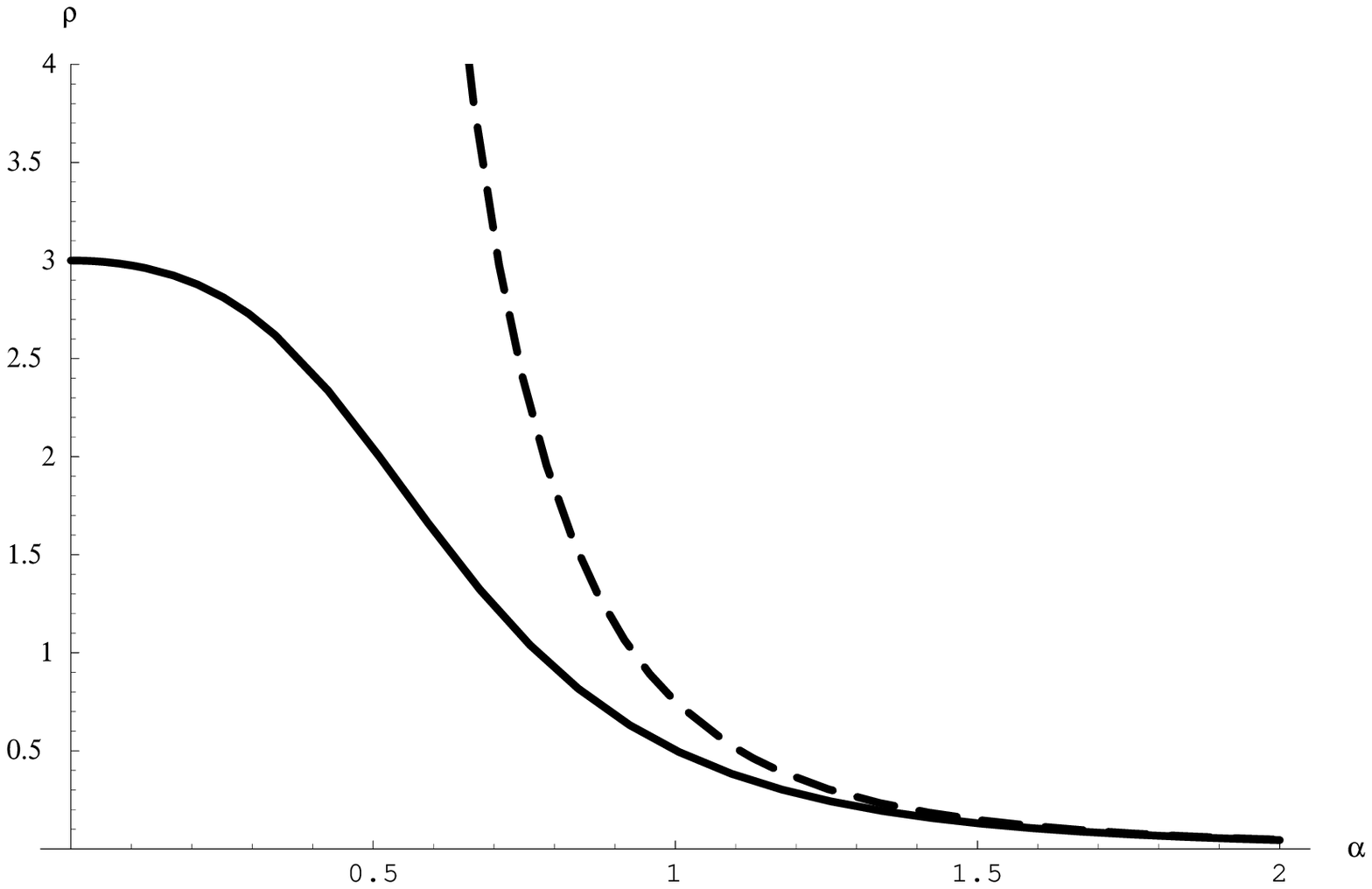}
\caption{$\rho(a) \times a$ for the case $k=0$, with $a_0 = \kappa = 1$,
$\alpha = 0.5$, $\beta = 0.01$. The
solid line uses the non-singular scale factor with inflationary behaviour;
the dashed line is obtained using the known scale factor, $a(t)=a_{0}\sqrt{t}$,
for radiation era in FRW cosmology.}
\label{figrho}
\end{center}
\end{figure}

The Figure \ref{figrho} shows the behaviour of $\rho (a)$, which has no
divergences if $\alpha >0$.

The pressure $p$ can be obtained from the Eq. (\ref{EnergyCons}), 
\begin{eqnarray}
p(a)&=&\frac{{a_{0}}^{4}\,\left( a^{2}+{a_{0}}^{2}\right) \,\left[
a^{6}+2\,a^{4}\,{a_{0}}^{2}+a^{2}\,{a_{0}}^{4}\,\left( 1-7\,\alpha \right)
-3\,{a_{0}}^{6}\,\alpha \right] }{4\,{\left( a^{4}+a^{2}\,{a_{0}}^{2}+{a_{0}}%
^{4}\,\alpha \right) }^{3}\,\kappa }+
\label{pk0} \\
& &\frac{9\,a^{2}\,{a_{0}}^{12}\,\left( a^{2}+{a_{0}}^{2}\right) \,\alpha
\,\,\beta }{2\,{\left( a^{4}+a^{2}\,{a_{0}}^{2}+{a_{0}}^{4}\,\alpha \right) }%
^{7}\,\kappa }\left\{ 126\,a^{12}+459\,a^{10}\,{a_{0}}^{2}+8\,a^{8}\,{a_{0}}%
^{4}\,\left( 85-36\,\alpha \right) +\right.
\nonumber \\
& &25\,{a_{0}}^{12}\,\left( -1+\alpha \right) \,\alpha -6\,a^{6}\,{a_{0}}%
^{6}\,\left( -87+119\,\alpha \right) +
\nonumber \\
& &\left. 3\,a^{4}\,{a_{0}}^{8}\left[\,70+\alpha \,\left( -207+22\,\alpha %
\right) \right] +a^{2}\,{a_{0}}^{10}\,\left[ 35+\alpha \,\left( -220+87\,%
\alpha\right) \right] \right\} \ ,
\nonumber
\end{eqnarray}

If $\alpha >0,$ the pressure $p$ is always negative for some region in the $%
a(t)$ or $t$ domain, independent of $\beta $. For $\alpha >0$, $p$ has no
divergences.

Testing whether $\rho +p\geq 0$ is true just restricts $\alpha $ and $%
\beta $ inside the region 
\begin{equation}
Q_{2}=\{0\leq \alpha <1,\beta _{-}<\beta <\beta _{+}\}\ ,
\label{Q2k0}
\end{equation}
where $\beta _{-}$ and $\beta _{+}$ are another functions of $\alpha $,
different from the analysis of $\rho $.

The velocity of sound, 
\begin{equation}
v_{s}^{2}=\frac{\partial p}{\partial \rho }=\frac{\partial p/\partial a}{%
\partial \rho /\partial a}\ ,
\label{v2k0}
\end{equation}
does not diverge if $\alpha $ and $\beta $ are inside the same region $Q_{2}$%
. But it can be imaginary for any value of $(\alpha ,\beta )\neq (0,0)$.

So the conclusion for the $k=0$ case is that the addition of high-order
terms in the Einstein action does not lead to any other anomaly in the
effective perfect fluid, i.e., with or without $\beta $, the behaviour of $%
\rho $, $p$, $\rho +p$ and $v_{s}$ is the same (under the condition that $%
\alpha $ and $\beta $ are inside the region $Q_{2}$).

\bigskip

{\bf II) The }$k=+1${\bf \ case :}

\medskip

\begin{figure}[h]
\begin{center}
\includegraphics[scale=0.80]{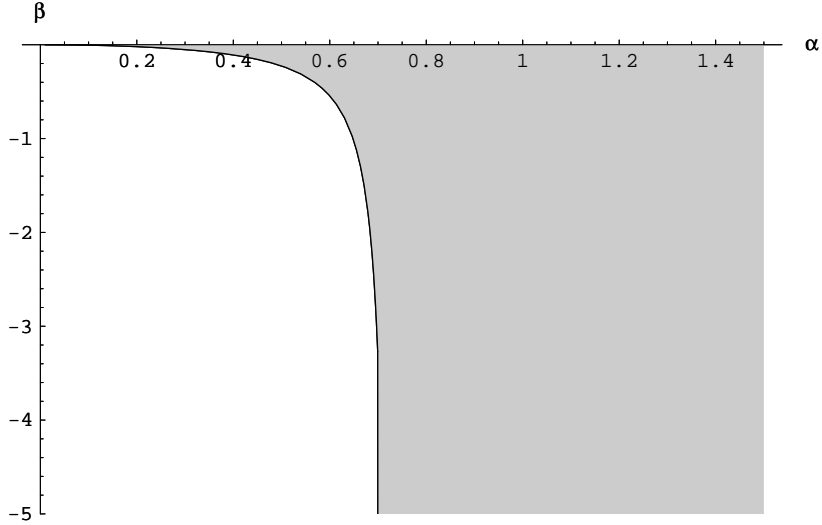}
\caption{The region $Q_1$ for the case $k=+1$, where $\rho > 0$, with
$a_0 = 1$.}
\label{figQ1}
\end{center}
\end{figure}

Analogous to the $k=0$ case, but with more difficult calculations, the
energy density is given by 
\begin{eqnarray}
\rho (a) &=&\frac{3\,{a_{0}}^{2}\,}{{\left( a^{5}+a^{3}\,{a_{0}}^{2}+2\,a\,{%
a_{0}}^{2}\,\sqrt{-a^{2}+{a_{0}}^{2}}\,\alpha \right) }^{2}\,\kappa }\left[
a^{6}+2\,a^{4}\,{a_{0}}^{2}+a^{2}\,{a_{0}}^{4}+\right.
\label{rhok1} \\
& &\left. 4\,a^{2}\,\sqrt{-a^{2}+{a_{0}}^{2}}\,\left( a^{2}+{a_{0}}^{2}\right)
\,\alpha -4\,a^{2}\,{a_{0}}^{2}\,{\alpha }^{2}+4\,{a_{0}}^{4}\,{\alpha }^{2}%
\right] +
\nonumber \\
& &\frac{216\,{a_{0}}^{2}\,\alpha \,\,\beta }{a^{4}\,{\left( a^{4}+a^{2}\,%
{a_{0}}^{2}+2\,{a_{0}}^{2}\,\sqrt{-a^{2}+{a_{0}}^{2}}\,\alpha \right) }%
^{6}\,\kappa }\left\{ a^{6}\,\left( a-a_{0}\right) \,{a_{0}}^{2}\,\left(
a+a_{0}\right) \times \right.
\nonumber \\
& &{\left( a^{2}+{a_{0}}^{2}\right) }^{2}\left( 51\,a^{6}+14\,a^{4}\,{a_{0}}%
^{2}-13\,a^{2}\,{a_{0}}^{4}+8\,{a_{0}}^{6}\right) \,\alpha + 4\,{a_{0}}^{6}\,%
{\left( a^{5}-a\,{a_{0}}^{4}\right) }^{2}\times
\nonumber \\
& &\left( -17\,a^{2}+2\,{a_{0}}^{2}\right) \,{\alpha }^{3}-16\,{a_{0}}^{10}\,%
{\left( -a^{2}+{a_{0}}^{2}\right) }^{3}\,{\alpha}^{5}+a^{2}\,\sqrt{-a^{2}+%
{a_{0}}^{2}} \times
\nonumber \\
& &\left( a^{4}-{a_{0}}^{4}\right)\,\left[a^{4}\,{\left( a^{2}+{a_{0}}^{2}%
\right) }^{2}\,\left( 5\,a^{6}-34\,a^{4}\,{a_{0}}^{2}-33\,a^{2}\,{a_{0}}^{4}%
-10\,{a_{0}}^{6}\right) +\right.
\nonumber \\
& &\left. \left.8\,a^{2}\,{a_{0}}^{4}\,\left( 7\,a^{2}-2\,{a_{0}}^{2}\right)%
\,{\left( a^{2}+{a_{0}}^{2}\right) }^{2}\,{\alpha }^{2}+48\,{a_{0}}^{8}\,%
\left( -a^{2}+{a_{0}}^{2}\right) \,{\alpha }^{4}\right] \right\} \ ,
\nonumber
\end{eqnarray}
and a long analysis shows that $\rho \geq 0$ if $\alpha $ and $\beta $
are inside the region 
\begin{equation}
Q_{1}=\{0\leq \alpha <\alpha _{\max },\beta _{-}<\beta \leq
0\}\cup \{\alpha >\alpha _{\max },\beta <0\}\ ,
\label{Q1k1}
\end{equation}
where $\beta _{-}$ is function of $\alpha $ and $a_{0}$, and $\alpha _{\max
} $ is a function of $a_{0}$, $\alpha _{\max }\approx 0.73\ a0$. Figure
\ref{figQ1} shows the region $Q_{1}$. The non-linear terms in the Lagrangian
do not force $\rho $ to be negative.

The pressure $p$ can be obtained from the Eq. (\ref{EnergyCons}) and will
not be shown here. Even $p\geq 0$ is possible if $\alpha $ and $\beta $
are inside the region 
\begin{equation}
Q_{2}=\{\alpha >\alpha _{\min },\beta <\beta _{-}\cup \beta >\beta
_{+}\}\ ,
\label{Q2k1}
\end{equation}
where $\beta _{-}$ and $\beta _{+}$ are functions of $\alpha $ and $a_{0}$,
and $\alpha _{\min }$ is a function of $a_{0}$, $\alpha _{\min }\approx
0.84\ a0$. As $\beta =0$ is not included in the region $Q_{2}$, this result
is a benefit of adding non-linear terms in the Lagrangian.

To obtain $\rho +p\geq 0$, $\alpha $ and $\beta $ must be inside the
region 
\begin{equation}
Q_{3}=\{0\leq \alpha <\alpha _{\max },\beta _{-}<\beta \leq
0\}\cup \{\alpha >\alpha _{\max },\beta <0\}\ ,
\label{Q3k1}
\end{equation}
where $\beta _{-}$ and $\beta _{+}$ are another functions of $\alpha $,
different from the analysis of $\rho $, and $\alpha _{\max }$ is a function
of $a_{0}$, $\alpha _{\max }\approx 0.80\ a0$. The regions $Q_{1}$ and $%
Q_{3} $ are similar.

The velocity of sound does not diverge if $\alpha $ and $\beta $ are inside
the same region $Q_{3}$. But it can be imaginary for any value of $(\alpha
,\beta )\neq (0,0)$; maybe only in the $Q_{2}$ region $v_{s}$ has real
values, this must be confirmed.

So the conclusion for the $k=+1$ case is that the addition of high-order
terms in the Einstein action does not lead to any other anomaly in the
effective perfect fluid, i.e., with or without $\beta $, the behaviour of $%
\rho $, $p$, $\rho +p$ and $v_{s}$ is the same (under the condition that $%
\alpha $ and $\beta $ are inside the region $Q_{3}$).

There is even an advantage because $p$ can be always positive, with only the
problem that the region $Q_{2}$ leads to almost suppression of the radiation
era. The calculus of the region $Q_{2}$ is not easy, and it is a task for
further work to verify if inflationary behaviour can exist with $\rho >0$, $%
p>0$ and $0<v_{s}<1.$

\bigskip

{\bf III) The }$k=-1${\bf \ case :}

\medskip

Like the $k=+1$ case, the calculations are very long, and the energy density
is given by 
\begin{eqnarray}
\rho (a)&=&\frac{3\,{a_{0}}^{2}\,\left( a^{2}+{a_{0}}^{2}\right) \,\left[
a^{4}-4\,{a_{0}}^{2}\,{\alpha }^{2}+a^{2}\,\left( {a_{0}}^{2}-4\,\sqrt{a^{2}+%
{a_{0}}^{2}}\,\alpha \right) \right] }{{\left( a^{5}+a^{3}\,{a_{0}}%
^{2}+2\,a\,{a_{0}}^{2}\,\sqrt{a^{2}+{a_{0}}^{2}}\,\alpha \right)}^{2}\,\kappa}+
\label{rhokm1} \\
& &\frac{216\,{a_{0}}^{2}\,\alpha \,\,\beta }{a^{4}\,{\left( a^{4}+a^{2}\,{
a_{0}}^{2}+2\,{a_{0}}^{2}\,\sqrt{a^{2}+{a_{0}}^{2}}\,\alpha \right)}^{6}\,%
\kappa }\left\{ -a^{6}\,{a_{0}}^{2}\,\left( 51\,a^{2}+32\,{a_{0}}^{2}\right)\,%
\right. \times
\nonumber \\
& &{\left( a^{2}+{a_{0}}^{2}\right)}^{5}\alpha -4\,a^{2}\,{a_{0}}^{6}\,{\left(
a^{2}+{a_{0}}^{2}\right) }^{4}\,\left(
17\,a^{2}+2\,{a_{0}}^{2}\right) \,{\alpha }^{3}-16\,{\left(a^{2}+{a_{0}}^{2}%
\right) }^{3} \times
\nonumber \\
& &{a_{0}}^{10}\,{\alpha }^{5}+a^{2}\,{\left(a^{2}%
+{a_{0}}^{2}\right) }^{7/2}\,\left[ a^{4}\,{\left( a^{2}+{a_{0}}^{2}%
\right) }^{2}\,\left( 5\,a^{4}+17\,a^{2}\,{a_{0}}^{2}+10\,{a_{0}}^{4}\right)%
-\right.
\nonumber \\
& &\left. \left. 8\,a^{2}\,{a_{0}}^{4}\,\left( a^{2}+{a_{0}}^{2}\right)
\,\left( 7\,a^{2}+2\,{a_{0}}^{2}\right) \,{\alpha }^{2}-48\,{a_{0}}^{8}\,{%
\alpha }^{4}\right] \right\} \ ,
\nonumber
\end{eqnarray}
and $\rho \geq 0$ is impossible for $\alpha >0$, independent of $\beta $%
.

The pressure $p$ will not be shown here. Like $\rho $, $p\geq 0$ is
impossible for $\alpha >0$ and any $\beta .$

Obviously, for $\alpha >0$ and any $\beta $ it is not possible to obtain $%
\rho +p\geq 0$.

The velocity of sound is divergent and imaginary if $\alpha >0$, independent
of $\beta $.

The case $k=-1$ does not provide useful conclusions, because even without
the high-order terms in the Einstein action there are a lot of anomalies in
the effective perfect fluid. The hope that quadratic-order terms could avoid
these anomalies was not achieved.

\section{Conclusion}

In all three cases of spatial curvature, it is possible to choose the
values of $\alpha$ and $\beta$ so that the quadratic-order terms in the
Lagrangian do not add more anomalies to the effective perfect fluid.

For closed spatial section, the opposite happens, i.e., anomalies are
suppressed, and this case of non-singular inflationary scale factor
with well-behaved perfect fluid deserves more investigation.

Other interesting improvements would be : use another solutions of the scale
factor, more energy conditions, compare the $\beta ^{\prime }s$ constraints
with observational values and study many types of effective perfect fluids.

\section*{ACKNOWLEDGMENTS}

I would like to thank Prof. Richard Kerner and the Laboratoire de
Gravitation et Cosmologie Relativistes of Universit\'{e} Pierre et Marie
Curie for proposing and supporting this work, and also CAPES of Brazil for
financial support.


\begin{thebibliography}{99}

\bibitem{Lanczos}  C. Lanczos, Ann. Math. {\bf 39}, 842 (1938).

\bibitem{Pais}  A. Pais and G. E. Uhlenbeck, Phys. Rev. {\bf 79}, 145 (1950).

\bibitem{DeWitt}  B. S. DeWitt, Phys. Rep. {\bf 19C}, 295 (1975).

\bibitem{Hooft}   G. 't Hooft and M. Veltman, Ann. Inst. Henri Poincar\'{e}
{\bf 20}, 69 (1974).

\bibitem{Stelle}  K. S. Stelle, Phys. Rev. D {\bf16}, 953 (1977).

\bibitem{Shapiro}  A. L. Maroto and I. L. Shapiro, Phys.Lett. B {\bf414},
34 (1997).

\bibitem{Starobinsky}  A. A. Starobinsky, Phys. Lett. B {\bf91}, 99 (1980).

\bibitem{Kerner}  J. P. Duruisseau and R. Kerner, Class. Quantum Grav.
{\bf 3}, 817 (1986).

\end{thebibliography}
\end{document}